\documentclass[english,twocolumn,showpacs,preprintnumbers,amsmath,amssymb,aps,prb,superscriptaddress,english]{revtex4}
\usepackage[T1]{fontenc}
\usepackage[latin9]{inputenc}
\usepackage{xcolor}
\usepackage{textcomp}
\usepackage{amstext}
\usepackage{graphicx}
\usepackage{amssymb}
\usepackage{esint}
\PassOptionsToPackage{normalem}{ulem}
\usepackage{ulem}
\usepackage{color}

\makeatletter

\providecommand{\tabularnewline}{\\}
\providecolor{lyxadded}{rgb}{0,0,1}
\providecolor{lyxdeleted}{rgb}{1,0,0}

\@ifundefined{textcolor}{}
{%
 \definecolor{BLACK}{gray}{0}
 \definecolor{WHITE}{gray}{1}
 \definecolor{RED}{rgb}{1,0,0}
 \definecolor{GREEN}{rgb}{0,1,0}
 \definecolor{BLUE}{rgb}{0,0,1}
 \definecolor{CYAN}{cmyk}{1,0,0,0}
 \definecolor{MAGENTA}{cmyk}{0,1,0,0}
 \definecolor{YELLOW}{cmyk}{0,0,1,0}
 }

\usepackage{graphicx}
\usepackage{dcolumn}
\usepackage{bm}

\let\oldAA=\AA
\renewcommand{\AA}{\ensuremath{\mbox{\oldAA}}}

\makeatother

\usepackage{babel}

\begin{document}

\title{Spinon heat transport and spin-phonon interaction in the antiferromagnetic
spin-1/2 Heisenberg chain cuprates Sr$_{2}$CuO$_{3}$ and SrCuO$_{2}$}

\author{N. Hlubek}

\affiliation{IFW-Dresden, Institute for Solid State Research, P.O. Box 270116,
D-01171 Dresden, Germany}

\author{X. Zotos}

\affiliation{Department of Physics and Institute of Theoretical and Computational
Physics, University of Crete, Greece}

\affiliation{Foundation for Research and Technology-Hellas, PO Box 2208, 71003
Heraklion, Greece}

\author{S. Singh}

\affiliation{IFW-Dresden, Institute for Solid State Research, P.O. Box 270116,
D-01171 Dresden, Germany}

\affiliation{Laboratoire de Physico-Chimie de L'Etat Solide, ICMMO, UMR8182, Universit\'{e}
Paris-Sud, 91405 Orsay, France}

\affiliation{Indian Institute of Science Education and Research, 900 NCL Innovation
Park, Pashan, Pune 411008, India}

\author{R. Saint-Martin}

\affiliation{Laboratoire de Physico-Chimie de L'Etat Solide, ICMMO, UMR8182, Universit\'{e}
Paris-Sud, 91405 Orsay, France}

\author{A. Revcolevschi}

\affiliation{Laboratoire de Physico-Chimie de L'Etat Solide, ICMMO, UMR8182, Universit\'{e}
Paris-Sud, 91405 Orsay, France}

\author{B. B\"{u}chner}

\author{C. Hess}

\affiliation{IFW-Dresden, Institute for Solid State Research, P.O. Box 270116,
D-01171 Dresden, Germany}

\date{\today}

\begin{abstract}
We have investigated the thermal conductivity $\kappa_{\mathrm{mag}}$
of high-purity single crystals of the spin chain compound Sr$_{2}$CuO$_{3}$
which is considered an excellent realization of the one-dimensional
spin-1/2 antiferromagnetic Heisenberg model. We find that the spinon
heat conductivity $\kappa_{\mathrm{mag}}$ is strongly enhanced as
compared to previous results obtained on samples with lower chemical
purity. The analysis of $\kappa_{\mathrm{mag}}$ allows to compute
the spinon mean free path $l_{\mathrm{mag}}$ as a function of temperature.
At low-temperature we find $l_{\mathrm{mag}}\sim0.5\,\text{\textmu m}$,
corresponding to more than 1200 chain unit cells. Upon increasing
the temperature, the mean free path decreases strongly and approaches
an exponential decay $\sim\frac{1}{T}\exp{T_{u}^{*}/T}$ which is
characteristic for umklapp processes with the energy scale $k_{B}T_{u}^{*}$.
Based on Matthiesen's rule we decompose {\normalsize $l_{\mathrm{mag}}$}
into a temperature-independent spinon-defect scattering length $l_{0}$
and a temperature dependent spinon-phonon scattering length $l_{\mathrm{sp}}(T)$.
By comparing {\normalsize $l_{\mathrm{mag}}(T)$} of Sr$_{2}$CuO$_{3}$
with that of SrCuO$_{2}$, we show that the spin-phonon interaction,
as expressed by {\normalsize $l_{\mathrm{sp}}$} is practically the
same in both systems. The comparison of the empirically derived {\normalsize $l_{\mathrm{sp}}$}
with model calculations for the spin-phonon interaction of the one-dimensional
spin-1/2 $XY$ model yields reasonable agreement with the experimental
data. 
\end{abstract}
\maketitle

\section{Introduction}

The physics of low-dimensional quantum magnets has recently attracted
considerable attention by experimentally and theoretically working
scientists because intriguing properties are found. Such systems exhibit
a variety of unusual ground states and exotic elementary excitations
which, in some cases are well accessible by theoretical treatments.
An important class of materials which host such quantum magnets is
formed by copper-oxides (cuprates) which feature Cu$^{2+}$-ions which,
through their 3$d^{9}$ configuration, generate $S=1/2$ sites. The
type and strength of the interaction between these spins depends crucially
on the structure of the material. For example, amongst the cuprates
there are model systems which realize $S=1/2$ arrangements with strong
antiferromagnetic (AFM) Heisenberg-type interaction ($J/k_{B}\sim2000\,$K)
in the form of square lattices, two-leg spin ladders and chains.\cite{Hayden91a,Thio88,Dagotto96,Dagotto1999,Motoyama1996}
{About 10 years} ago it was discovered, that in all these different
systems their magnetic excitations give rise to a highly anisotropic
thermal conductivity tensor of the respective materials, with an unexpectedly
large magnetic contribution along the directions of large AFM exchange.
This finding opened up a new route to investigating the generation,
scattering and dissipation of magnetic quasiparticles (complementary
to neutron scattering and magnetic resonance experiments) through
analyzing the magnetic thermal conductivity $\kappa_{\mathrm{mag}}$.\cite{Sologubenko2000a,Sologubenko2000b,Sologubenko2001,Hess2001,Hess2004a,Hess2005,Hess2006,Hess2003a,Hess2007a,Hlubek2010,Hlubek2011}
Among these findings, the results for compounds which realize the
one-dimensional $S=1/2$ AFM Heisenberg model (1D-AFM-HM) are particularly
interesting because fundamental conservation laws predict
ballistic heat transport in these systems.\cite{Zotos1997,Zotos1999}
This means that in the 1D-AFM-HM model, an \emph{infinite} magnetic heat
conductivity is expected. Despite this rigorous prediction, in any
real system, the transport is dissipative, due to extrinsic scattering
mechanisms. Nevertheless, an unprecedentedly large $\kappa_{\mathrm{mag}}$
has been observed in ultra-pure samples of the zig-zag chain compound
SrCuO$_{2}$.\cite{Hlubek2010,Kawamata2010} More specifically, upon
enhancing the chemical purity of the compound, $\kappa_{\mathrm{mag}}$
becomes i) increasingly larger, and ii) even for the
highest purity level it can be fully described (in the framework of
a simple kinetic model) by considering spinons scattering off impurities
and phonons only, where the impurity scattering fully accounts for
purity dependence of $\kappa_{\mathrm{mag}}$ and the phonon scattering
prevails at elevated temperatures. This is indeed consistent with
the predicted ballistic transport since no further scattering (i.e.
spinon spinon scattering) mechanism needs to be invoked.\cite{Hlubek2010}
Furthermore, the analysis of the data yields very clean data for the
spinon-phonon scattering, for which a full theoretical description
is still lacking.

In the zig-zag chain compound SrCuO$_{2}$, two $S=1/2$ chains with
large AFM exchange are tied together by a significant but frustrated
intrachain exchange. In this paper we extend the previous findings for SrCuO$_{2}$ to the
\emph{single-}chain material Sr$_{2}$CuO$_{3}$, where such complications
are absent. We find that in high purity samples of this compound $\kappa_{\mathrm{mag}}$
is strongly enhanced as compared to previous results\cite{Sologubenko2001,Kawamata2008}
for lower purity. Upon comparing the dependence of $\kappa_{\mathrm{mag}}$
on the nominal purity level we observe that $\kappa_{\mathrm{mag}}$
of Sr$_{2}$CuO$_{3}$ depends in a similar fashion on the purity
of the material like SrCuO$_{2}$. More specifically, within
a {simple kinetic model\cite{Sologubenko2001}} the spinon mean free path $l_{\mathrm{mag}}$
can be decomposed into terms which describe spinon-defect and spinon-phonon
scattering as is the case for the zig-zag chain compound. We show
that using an {empirical formula\cite{Sologubenko2001}} for describing the spinon-phonon scattering
that the strength of this mechanism is practically indistinguishable
for both materials. Going beyond this empirical approach we model
the spinon-phonon scattering by employing results for the spin-phonon
interaction of the $XY$-model which further underpins these findings.

\section{Materials Details}

The main building blocks of Sr$_{2}$CuO$_{3}$ are corner sharing
chains formed by CuO$_{3}$ plaquettes.\cite{Teske1969a} The chains
are parallel to the crystallographic $b$-axis in Sr$_{2}$CuO$_{3}$.
The intrachain exchange interaction between neighboring Cu$^{2+}$
sites is mediated by 180\textdegree{} superexchange through the oxygen
and with $J/k_{B}\approx2150-3000$\,K\cite{Ami1995,Suzuura1996,Motoyama1996}
is among the largest among known Heisenberg spin chain materials.
The Cu-O-Cu chains are separated by Sr atoms, leading to an extremely
small interchain interaction of $J_{\mathrm{perp}}/k_{B}\approx0.02$\,K.
Furthermore, muon spin rotation and neutron scattering measurements\cite{Keren1993,Kojima1997}
show that this material orders three-dimensionally only below the
N\'{e}el temperature of $T_{N}\approx5.4$\,K. These properties make
Sr$_{2}$CuO$_{3}$ an excellent realization of a $S=1/2$ Heisenberg
chain for temperatures $T>5.4$\,K. While Sr$_{2}$CuO$_{3}$ consists
of isolated chains, the main structural element in $\mathrm{SrCuO_{2}}$
is formed by $\mathrm{CuO_{2}}$ zig-zag ribbons, which run along
the crystallographic $c$-axis. Each ribbon can be viewed as made
of two parallel chains of corner-sharing $\mathrm{CuO_{2}}$ plaquettes,
where the straight Cu-O-Cu bonds between corner-sharing plaquettes
of each double-chain structure result in a very large antiferromagnetic
intrachain exchange coupling $J/k_{B}\approx2100-2600$ K of the $S=1/2$
spins at the $\mathrm{Cu^{2+}}$sites.\cite{Zaliznyak2004,Motoyama1996}
The interchain coupling $J^{'}$ between Cu$^{2+}$-sites of two edge-sharing
plaquettes is much weaker ($\left|J^{'}\right|/J\approx0.1-0.2$).\cite{Rice1993,Motoyama1996}
Frustration of this exchange interaction and presumably quantum fluctuations
prevent three-dimensional long range magnetic order of the system
at $T>T_{N}\approx1.5-2\,\mathrm{K}\approx10^{-3}J/k_{B}$~K.\cite{Matsuda1997a,Zaliznyak1999}
Hence, at significantly higher $T$ the two chains within one double
chain structure are usually regarded as magnetically independent.
In fact, low-$T$ (12~K) inelastic neutron scattering spectra of
the magnetic excitations can be very well described within the $S=1/2$
Heisenberg antiferromagnetic chain model.\cite{Zaliznyak2004}

\section{Experimental details }

Large single crystals of pure $\mathrm{Sr_{\mathrm{2}}CuO_{3}}$ were
grown by the traveling solvent floating zone method.\cite{Revcolevschi1999}
The feed rods were prepared using the primary chemicals $\mathrm{CuO}$
and $\mathrm{SrCO_{3}}$ with 4N (99.99\%) purity. The crystals react
rather rapidly with water. A hydroxide layer is formed at the surface
of the crystals, after a short exposure to air.\cite{Scholder1933,Hill2002}
Therefore the crystals were annealed at 900\textdegree{}C in an oxygen
atmosphere for three days before further treatment. This reverts the
hydroxide layer back to Sr$_{2}$CuO$_{3}$, although in a polycrystalline
form. 

The crystallinity and stoichiometry of all crystals were checked under
polarized light and by energy-dispersive X-ray spectroscopy, respectively.
For the transport measurements rectangular samples with typical dimensions
of $\left(2\cdot0.5\cdot0.5\right)\,\mbox{mm}^{3}$ were cut from
the crystals for each doping level with an abrasive slurry wire saw.
Four-probe measurements of the thermal conductivity $\kappa$ were
performed\cite{Hess2003b} in the 7\textendash{}300\,K range with
the thermal current along the $b$ and $c$-axes ($\kappa_{b}$ and
$\kappa_{c}$ respectively).

\section{Experimental Results}

\begin{figure}
\begin{centering}
\includegraphics[width=8.5cm]{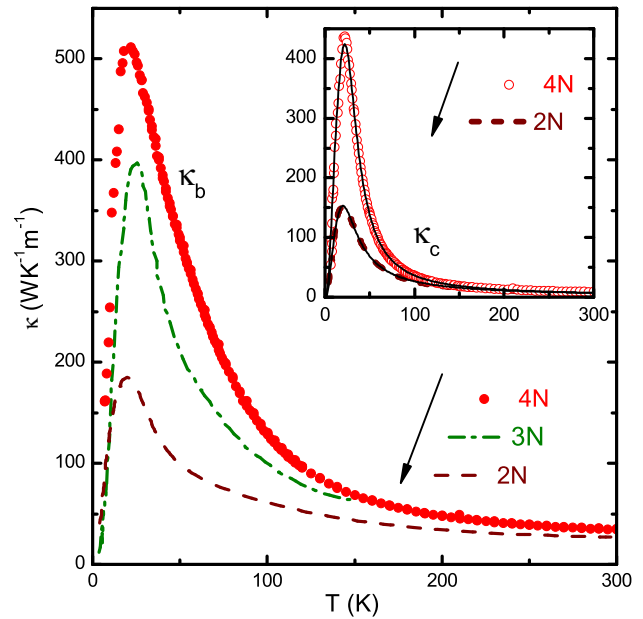}
\par\end{centering}

\caption{Thermal conductivity of Sr$_{2}$CuO$_{3}$ parallel to the spin chains
along $\kappa_{b}$ for various purities. The dashed lines represent
results from Sologubenko et. al. with 2N purity, reproduced from Ref.~\onlinecite{Sologubenko2000b}.
The dash-dotted line has been obtained by Kawamata et. al. for 3N
purity and is reproduced from Ref.~\onlinecite{Kawamata2008}. Inset:
Thermal conductivity of Sr$_{2}$CuO$_{3}$ perpendicular to the spin
chains along $\kappa_{c}$ for 2N (also reproduced from Ref.~\onlinecite{Sologubenko2000b})
and 4N purity. The solid lines are fits to the Callaway model. \label{fig:kappa of Sr2CuO3}}

\end{figure}

Fig.~\ref{fig:kappa of Sr2CuO3} presents our findings for the heat
conductivity of Sr$_{2}$CuO$_{3}$, measured with the heat current
parallel to the $b$ and $c$ axes, i.e. parallel and perpendicular
to the chains in the material. We focus first on the temperature dependence
of the thermal conductivity perpendicular to the spin chain, $\kappa_{c}$,
which is shown in the inset of the figure. Along this direction, the
heat conductivity of this electrically insulating material is purely
phononic: As a function of temperature, it shows a characteristic
peak at $T=22$\,K, and then strongly decreases upon further rising
the temperature. The height of the peak sensitively depends on the
density of impurities in the system, which generate phonon-defect
scattering. This can be well inferred by comparing our data for a
4N-purity material with that of 2N (i.e., 99\%) purity, taken from
Ref.~\onlinecite{Sologubenko2001}. For this lower-purity sample
the overall magnitude of $\kappa_{c}$ is strongly reduced as is expected
for typical phonon heat conductors.\cite{Berman} In fact, the data
for both purities can be well described in the framework of a model
by Callaway\cite{Callaway1959}, where the difference between both
curves is largely captured by different point defect scattering strength
(see Appendix).

The thermal conductivity parallel to the chain, $\kappa_{b}$, is
shown in the main panel of Fig.~\ref{fig:kappa of Sr2CuO3}. $\kappa_{b}$
exhibts a peak at the same position as observed for $\kappa_{c}$.
However, the peak is much broader and the overall magnitude of $\kappa_{b}$
is significantly larger than that of the purely phononic $\kappa_{c}$,
which is the signature of a substantial magnetic contribution in this
material, i.e., the heat current parallel to the spin chain is carried
not only by phonons but also by spinons.\cite{Sologubenko2001} For
our 4N purity sample, the anisotropy between $\kappa_{b}$ and $\kappa_{c}$
is roughly constant above 100\,K and approximately $\kappa_{b}/\kappa_{c}\approx3.5$.
This is significantly larger than the previously reported\cite{Sologubenko2001}
anisotropy for 2N-purity Sr$_{2}$CuO$_{3}$ and provides clear evidence
that the enhanced purity leads to a relative enhancement of the spinon
contribution to the overall heat conductivity. The purity dependence
of $\kappa_{b}$ can directly be read off from the figure where we compare
our findings with experimental data for 3N (99.9\% purity) and 2N
purity samples of Sr$_{2}$CuO$_{3}$, reproduced from Ref.~\onlinecite{Kawamata2008}
and Ref.~\onlinecite{Sologubenko2001}, respectively. From low to
intermediate temperatures $\left[7\mathrm{\, K}-150\mathrm{\, K}\right]$,
the heat conductivity is strongly enhanced upon increasing the sample
purity. At higher temperatures this purity dependence becomes weaker
since the curves approach each other. Apparently it is possible to
separate the thermal conductivity into two distinct regimes, where
different scattering processes dominate. At low-$T$, the extreme
sensitivity to impurities suggests, that spinon scattering off defects
is dominating. At high-$T$, a further extrinsic scattering mechanism
-- {spinon-phonon} scattering\cite{Sologubenko2001} -- becomes dominating
as a consequence of increasing phonon population, which leads to the
very similar $\kappa_{b,\mathrm{2N}}$ and $\kappa_{b,\mathrm{4N}}$
for $T\gtrsim200$\,K.

\section{Data analysis and discussion}

\subsection{The spinon heat conductivity of Sr$_{2}$CuO$_{3}$}

\begin{figure}
\begin{centering}
\includegraphics[bb=0bp 0bp 241bp 241bp,width=8.5cm]{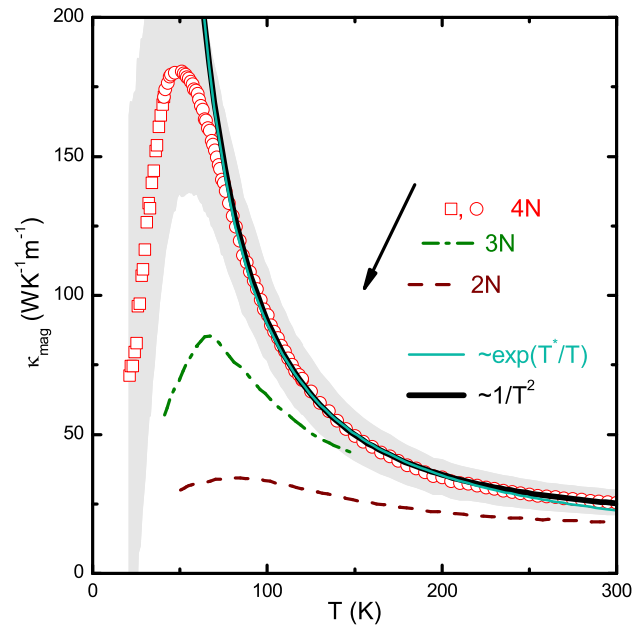}
\par\end{centering}

\caption{Estimated magnetic thermal conductivity of Sr$_{2}$CuO$_{3}$ for
4N (circles, squares), 3N (dash-dotted line)~\cite{Kawamata2008},
and 2N (dashed line)~\cite{Sologubenko2000b} purity. The shaded
area illustrates the uncertainty from the estimation of the phononic
background. The 4N results shown in squares instead of circles have
a large uncertainty. The thick solid line is a fit for $T>80\,\mathrm{K}$
with $\kappa_{\mathrm{mag,fit1}}\sim1/T^{2}$. The thin solid line
is a fit with $\kappa_{\mathrm{mag,fit2}}\sim\exp\left(T^{*}/T\right)$.
\label{fig:kappa mag of Sr2CuO3}}

\end{figure}

The thermal conductivity parallel to the chain is composed of a magnetic
and phononic contribution $\kappa_{b}=\kappa_{\mathrm{mag}}+\kappa_{b,\mathrm{ph}}$.
In order to extract the heat conductivity of the spin chain the phononic
background is approximated as $\kappa_{b,\mathrm{ph}}\approx\kappa_{c}$.
This simple assumption is reasonable, since the purely phononic anisotropy
between $\kappa_{a}$ and $\kappa_{c}$ {is small.\cite{Sologubenko2000b,Sologubenko2001}}
Therefore $\kappa_{b,\mathrm{ph}}$ is not expected to be much different.
The thus obtained spinon heat conductivity $\kappa_{\mathrm{mag}}=\kappa_{b}-\kappa_{c}$
for the 4N sample, as well as the results from literature for 3N and
2N, {for which $\kappa_{\mathrm{mag}}$ was extracted in a similar way,\cite{Sologubenko2000b,Sologubenko2001,Kawamata2008}} are shown in Fig.~\ref{fig:kappa mag of Sr2CuO3}. Below $T\lesssim40$\,K,
i.e., in the vicinity of the peak of $\kappa_{\mathrm{ph},b}$, errors
become large and the data in this range are neglected for further analysis.
For higher $T$, the possible uncertainty of $\kappa_{\mathrm{mag}}$
is around $\pm15$\%, which accounts for the individual errors of
$\kappa_{b}$ and $\kappa_{c}$. 

Starting from low-$T$, $\kappa_{\mathrm{mag}}$ of the 4N sample
increases almost linearly towards a peak. The uncertainty is quite
large in the temperature regime, and hence an increase with a higher
power\cite{Rozhkov2005,Chernyshev2005} of $T$ in this regime ($T<40$~K)
cannot be excluded. However, in the simplest case of a temperature
independent spinon-defect scattering rate, one expects $\kappa_{\mathrm{mag}}$
to be directly proportional to the thermal Drude weight $D_{\mathrm{th}}$,
which also increases linearly with temperature at low-$T$ up to $T\sim0.15J/k_{B}\sim300$~K.\cite{Kluemper2002,Heidrich02,Heidrich2003,Heidrich2005,Hess2007a}
The peak is quite pronounced and found at $\sim$48\,K with a maximum
value of about $180\mathrm{\, Wm^{-1}K^{-1}}$. The peak is followed
by a strong decrease for higher $T$. Such a temperature dependence
can not be accounted for from the $T$-dependence of the thermal Drude
weight, since it is expected to decrease only at very high $T\gtrsim0.6J/k_{B}\sim1200$~K.\cite{Kluemper2002,Heidrich02,Heidrich2003,Heidrich2005}
Instead, the decrease is consistent with our earlier notion of dominant
spinon-phonon scattering at high-$T$. $\kappa_{\mathrm{mag}}$ of
the 3N and 2N samples is qualitatively very similar to that of the
4N sample but is increasingly suppressed with growing impurity level,
accompanied with a shift of the peak-position of $\kappa_{\mathrm{mag}}$
towards higher temperatures. Furthermore, the curves approach each
other with increasing temperature, as the $\kappa_{b}$ data.

Chernyshev and Rozhkov have proposed a model which describes $\kappa_{\mathrm{mag}}$
of Sr$_{2}$CuO$_{3}$ at the 2N purity level very well.\cite{Chernyshev2005,Rozhkov2005}
However, a similarly convincing description of our data for the 4N
sample is not possible. The same holds for the double chain material
SrCuO$_{2}$. Spin-phonon drag has been suggested as a possible explanation
for the failure of the model.\cite{Chernyshev2007} Recently there
has been substantial progress in the formal theoretical treatment
of this phenomenon.\cite{Gangadharaiah2010} However, specifi{}c
model calculations have not yet been performed for the materials under
scrutiny here. It thus remains unclear whether spin-phonon drag plays
a significant role in our experimental data, and we analyse $\kappa_{\mathrm{mag}}$
without taking into account a possible contribution due to this effect. 

The temperature dependence of $\kappa_{\mathrm{mag}}$ at high temperature
$T\gtrsim80$~K up to room temperature can equally well be described
by either $\kappa_{\mathrm{mag}}\propto1/T^{2}+\mathrm{const.}$ or
$\kappa_{\mathrm{mag}}\propto\exp(T_{u}^{*}/T)$, with $T_{u}^{*}$
a characteristic energy scale. Fig.~\ref{fig:kappa mag of Sr2CuO3}
shows the corresponding fits. While the physical meaning of the former
functional form remains elusive, one expects the exponential one for
spinons scattering off phonons from general considerations for Umklapp
processes.\cite{AshcroftMermin1976} 

\begin{figure}
\begin{centering}
\includegraphics[width=8.5cm]{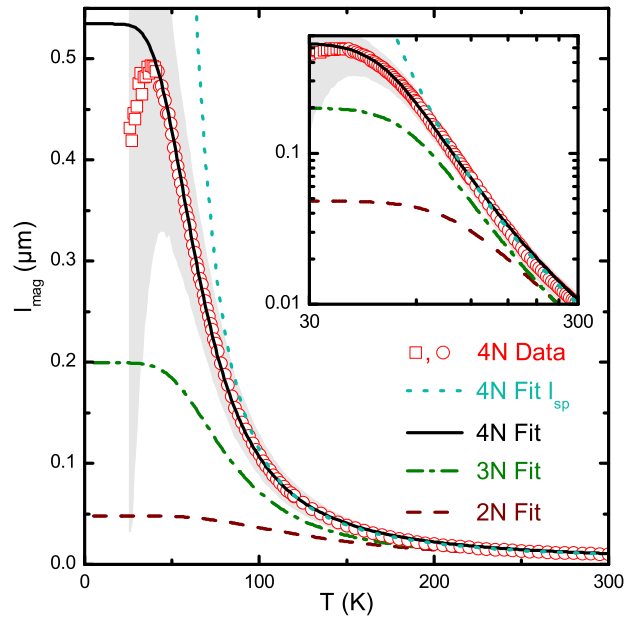}
\par\end{centering}

\caption{Derived mean free paths of the spinon excitations in Sr$_{2}$CuO$_{3}$
for 4N (circles, squares) purity. The solid black line is a fit to
the data as explained in the text. The squares in the 4N data have
a large uncertainty and are disregarded for the fit. A comparison
is made with the fits for 2N (dashed line)\cite{Sologubenko2001}, and
3N (dash-dotted line)\cite{Kawamata2008} purities. The dotted line
is a fit done to the 4N results by $l_{\mathrm{sp}}$ (eq.~\ref{eq:general Umklapp})
only. The shaded area illustrates the uncertainty from the estimation
of the phononic background. The inset shows the same results in a
double logarithmic scale. \label{fig:l mag of Sr2CuO3}}

\end{figure}

\subsection{The spinon mean free path}

We proceed by calculating the mean free path of the spinons, $l_{\mathrm{mag}}$,
from the {experimental $\kappa_{\mathrm{mag}}$ by\cite{Hess2007a,Sologubenko2000b,Sologubenko2001,Hess2007b,Hlubek2010}}
\begin{equation}
l_{\mathrm{mag}}=\frac{3\hbar}{\pi N_{s}k_{B}^{2}T}\kappa_{\mathrm{mag}},\label{eq:l mag data}\end{equation}
where $N_{s}$ is the number of spin chains per unit area. The thus
extracted mean free paths are shown in Fig.~\ref{fig:l mag of Sr2CuO3}.
$l_{\mathrm{mag}}(T)$ is very large at low temperature ($\sim0.5$~\textmu{}m)
and decreases strongly with increasing temperature, consistent with
the above already inferred increasing importance of spinon-phonon
scattering. While the high temperature data ($T\gtrsim100$~K) reflects
well the exponential suppression of $\kappa_{\mathrm{mag}}$, i.e.
$l_{\mathrm{mag}}\sim\frac{1}{T}\exp{T_{u}^{*}/T}$, (shown as a dotted
line in Fig.~\ref{fig:l mag of Sr2CuO3}), the low temperature data
clearly deviate from this functional form, which indicates that spinon-defect scattering becomes
important. In order to test this notion, we compare in Fig.~\ref{fig:l mag of Sr2CuO3}
these $l_{\mathrm{mag}}$ data for the 4N sample with fits to the
results of 2N and 3N purity samples as given in Refs.~\onlinecite{Sologubenko2001, Kawamata2008}.
As expected, the enhanced impurity density in these samples causes
a corresponding reduction of $l_{\mathrm{mag}}$ at low temperature.
Note that at high temperature the curves cling to that of the 4N sample,
which corresponds to the natural expectation of an identical spinon-phonon
scattering strength in all samples.

In order to capture this behavior in our further analysis, we apply
Matthiessen's rule for the scattering processes of the spinons $l_{\mathrm{mag}}^{-1}\left(T\right)=l_{0}^{-1}+l_{\mathrm{sp}}^{-1}\left(T\right)$,
where $l_{0}$ denotes the $T$-independent spinon-defect scattering,
while $l_{\mathrm{sp}}\left(T\right)$ takes the $T$-dependent spinon-phonon
scattering into account. According to our empirical finding of an
exponential decay of $\kappa_{\mathrm{mag}}(T)$ at high temperature
and consistent with previous findings\cite{Sologubenko2001,Hlubek2010}
we estimate $l_{\mathrm{sp}}\left(T\right)$ as

\begin{equation}
l_{\mathrm{sp}}^{-1}=\left(\frac{\exp\left(T_{u}^{*}/T\right)}{A_{s}T}\right)^{-1},\label{eq:general Umklapp}\end{equation}
with fit parameters $T_{u}^{*}$ and $A_{s}.$ As can be seen in Fig.~\ref{fig:l mag of Sr2CuO3}
such a empirically derived functional form for the spinon-phonon scattering
allows an excellent fit of the data with $l_{0}=0.54\pm0.05$~\textmu{}m
(corresponding to approximately 1367 lattice spacings), $T_{u}^{*}=210\pm11$~K,
and $A_{s}=(6.1\pm5)\times10^{5}\mathrm{m^{-1}K^{-1}}$.

\subsection{Comparison of SrCuO$_{2}$ and Sr$_{2}$CuO$_{3}$}

\begin{figure}
\begin{centering}
\includegraphics[width=8.5cm]{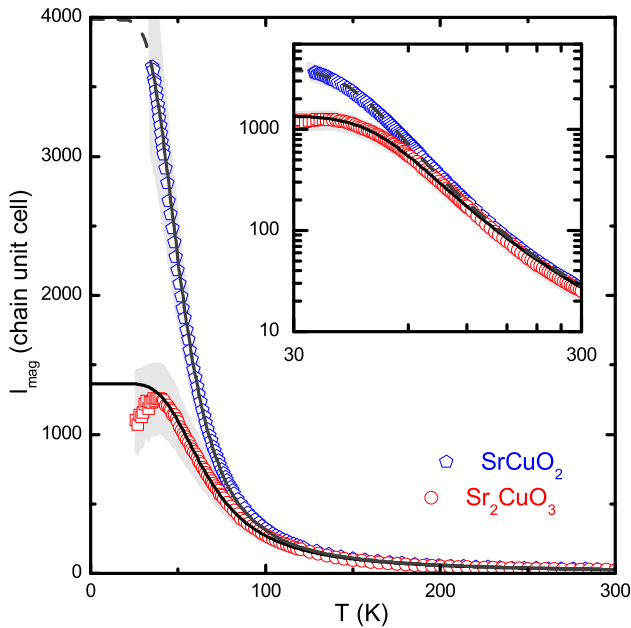}
\par\end{centering}

\caption{Magnetic mean free paths of SrCuO$_{2}$ (pentagon shape) and Sr$_{2}$CuO$_{3}$
(circles) for 4N purity. The shaded area illustrates the uncertainty
from the estimation of the phononic background. The lines are fits
to the mean free paths. \label{fig:SrCuO2 - Sr2CuO3 - Comparison}}

\end{figure}

The drastic enhancement of $\kappa_{\mathrm{mag}}$ and $l_{\mathrm{mag}}$
upon increasing the purity level of Sr$_{2}$CuO$_{3}$ provides strong
evidence that the spinon heat transport in the $S=1/2$ AFM Heisenberg
model as realized in this material is only limited by the \emph{external}
scattering off defects and phonons. Our findings here thus further
corroborate previous experimental evidence\cite{Hlubek2010} for
the ballistic nature of heat transport in the $S=1/2$ AFM Heisenberg
model, obtained for the zig-zag chain compound SrCuO$_{2}$. Since
the individual chains in SrCuO$_{2}$ and in Sr$_{2}$CuO$_{3}$ consists
of the same structural elements (CuO$_{2}$ plaquettes), it is instructive
to directly compare the findings for spinon-phonon scattering obtained
for both compounds. For this purpose we show in Fig.~\ref{fig:SrCuO2 - Sr2CuO3 - Comparison}
the spinon mean free path $l_{\mathrm{mag}}(T)$ of high-purity (4N)
SrCuO$_{2}$ and Sr$_{2}$CuO$_{3}$ in units of the chain unit cells.
In this representation, the mean free path is free of any geometrical
particularities and can directly be related to distances of spin sites
within a single chain. As can be seen in the figure, the mean free
paths for both compounds are virtually identical for temperatures
$T>150$\,K (the relative difference is less than 5\%). In the whole
temperature range, both curves can be well fitted with expression~\ref{eq:general Umklapp}
for the spinon-phonon scattering, with the same $T_{u}^{*}$, $A_{s}$
as determined afore and with and $l_{0}=1.56\pm0.16$~\textmu{}m
(corresponding to approximately 3984 lattice spacings) for SrCuO$_{2}$.
In fact, the same parameters for $l_{\mathrm{sp}}\left(T\right)$
could be used for both compounds. This demonstrates that the spinon
phonon interaction is the same in both compounds,
despite the difference of their CuO$_{2}$ chain structures. Since
the chains in SrCuO$_{2}$ and Sr$_{2}$CuO$_{3}$ are composed of
the same copper-oxygen plaquettes, one has to conclude that only phonon
modes which modulate the Cu-O-Cu bonds along two corner-sharing plaquettes
lead to a significant scattering of spinons.

At low temperature, the mean free path of SrCuO$_{2}$ is by a factor
of three larger than that of Sr$_{2}$CuO$_{3}$, in spite of the
same nominal purity. The higher defect density of Sr$_{2}$CuO$_{3}$
may be caused by the relatively lower chemical stability. It reacts
quite rapidly with water and decays, if exposed to air for a few
hours. Additionally, with regard to unavoidable intrinsic crystal
defects, the double chain is a much more stable structure,
due to its layout of the CuO$_{2}$-plaquettes. 

Apart from these minor and plausible differences with regard to the
spinon heat conduction, we would like to point out a surprising dissimilarity
which concerns the phonon heat conductivity $\kappa_{\mathrm{ph}}$.
As we have seen in Fig.~\ref{fig:kappa of Sr2CuO3}, $\kappa_{\mathrm{ph}}$
of Sr$_{2}$CuO$_{3}$ increases strongly at increasing the purity,
as expected. However, in SrCuO$_{2}$ the increase is very weak, which
suggests an additional scattering mechanism for phonons in this
material.\cite{Hlubek2010}

\subsection{Theoretical treatment of the spinon-phonon scattering}

\begin{figure}
\begin{centering}
\includegraphics[width=8.5cm]{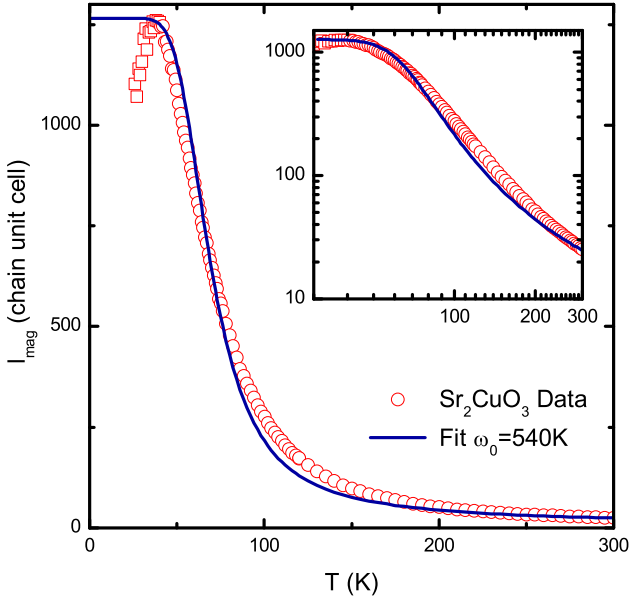}
\par\end{centering}

\caption{Fits to the mean free path according to the memory function approach
explained in the text. The circles are the derived mean free paths
of the spinon excitations in Sr$_{2}$CuO$_{3}$ for 4N purity. The
line represent a fit using $\tilde{l}_{\mathrm{sp}}$. The parameters
of the fit are $\omega_{0}=540\,\mathrm{K}$ and $l_{0}=1266$~lattice
spacings ($l_{0}=4955\,\mathrm{\AA}$). \label{fig:l fit of Sr2CuO3 with MM}}

\end{figure}

\begin{figure}
\begin{centering}
\includegraphics[width=8.5cm]{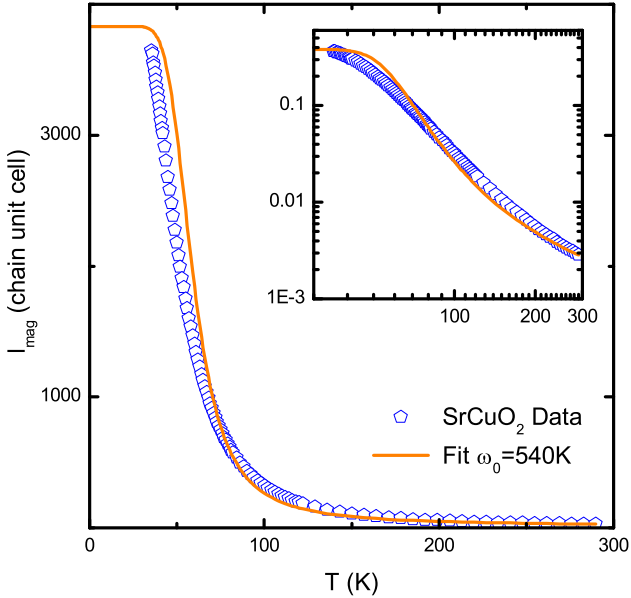}
\par\end{centering}

\caption{Fits to the mean free path according to the memory function approach
explained in the text. The pentagonal shaped symbols represent the
derived mean free paths of the spinon excitations in SrCuO$_{2}$
for 4N purity. The line represents a fit using an $l_{\mathrm{sp}}$
as defined by equation~\ref{eq:mm l_sp}. The parameters of the fit
are $\omega_{0}=540\,\mathrm{K}$ and $l_{0}=3831$~lattice spacings
($l_{0}=15000\,\mathrm{\AA}$). \label{fig:l fit of SrCuO2 with MM}}

\end{figure}

\begin{figure}
\begin{centering}
\includegraphics[width=8.5cm]{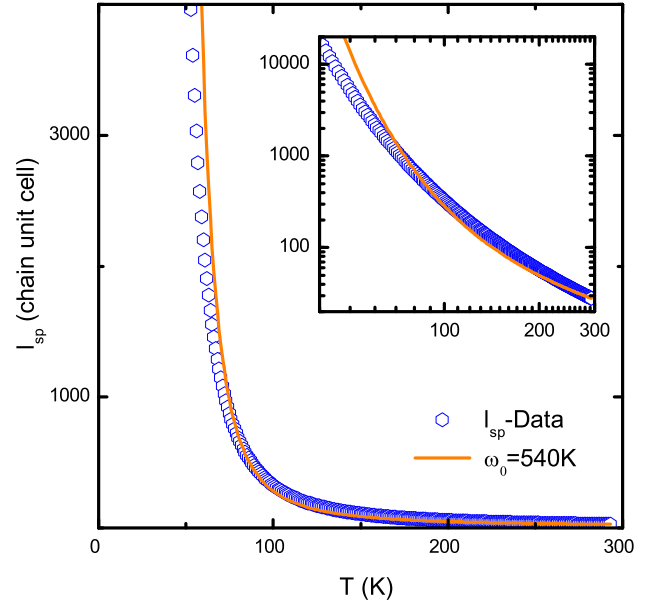}
\par\end{centering}

\caption{Comparison of the expressions for $l_{\mathrm{sp}}$ only. The hexagonal
shaped symbols represent an estimate for $l_{\mathrm{sp}}$ derived
from the magnetic mean free path $l_{\mathrm{mag}}$ according
to $l_{\mathrm{sp}}^{-1}\left(T\right)=l_{\mathrm{mag}}^{-1}\left(T\right)-l_{0}^{-1}$.
An $l_{0}=1.56\pm0.16$~\textmu{}m as approximated by the phenomenological
model has been used. The line represents calculations of $l_{\mathrm{sp}}$
with the memory function approach using a phonon frequency $\omega_{0}=540\,\mathrm{K}$.
\label{fig:l sp comparison - Data vs MM}}

\end{figure}

Expression~\ref{eq:general Umklapp} which describes the spinon-phonon
scattering in our data has been empirically derived and is consistent
with general considerations for Umklapp scattering. In the following
we go beyond this empirical treatment and derive $l_{\mathrm{sp}}\left(T\right)$
from a spinon-phonon scattering theory presented in Ref.~\onlinecite{Louis2006}
(labelled $\tilde{l}_{\mathrm{sp}}(T)$). In this memory function
approach it is possible to semi-analytically evaluate the temperature
dependence of the mean-free path within the XY limit of the Heisenberg
model assuming weak coupling.

Applying a Jordan-Wigner transformation the XY Hamiltonian becomes,
\begin{eqnarray}
H=\sum_{l}h_{l,l+1}^{s}=-t\sum_{l}(1-\lambda(x_{l+1}-x_{l}))(c_{l+1}^{\dagger}c_{l}+h.c.)\end{eqnarray}

\noindent where $-t=J/2$ and $\lambda$ is the spin-phonon coupling
constant in the spin-Peierls XY model. In this tight binding model
the dispersion of fermions is given by $\epsilon_{k}=-2t\cos(ka)$
and the velocity at the Fermi wavevector $k=\pi/2a$ is equal to $v=2ta/\hbar$
($a$ is the lattice constant). In the isotropic Heisenberg model
the elementary magnetic excitations -- the spinons -- have a velocity
equal to $v_{\mathrm{sp}}=\frac{\pi}{2}\frac{aJ}{\hbar}$. To take
into account the difference in velocities of the excitations between
the XY and isotropic Heisenberg model we evaluate the mean free path
using an effective hopping matrix element $-t_{\mathrm{eff}}=\frac{\pi}{2}\frac{J}{2}$.

The mean free path is given by $\tilde{l}_{\mathrm{sp}}\left(T\right)\sim v_{\mathrm{sp}}\tau\left(T\right)$
where $\tau$ is the characteristic scattering time. The phase space
of the spinon-phonon scattering matrix elements determining $\tau$
is restricted by the energy-momentum conservation laws\cite{Louis2006}
as, \begin{eqnarray}
1/\tau & \sim & f_{k}(1-f_{k+q})[(1+n_{-q})\delta(\hbar\omega+\epsilon_{k}-\epsilon_{k+q}-\hbar\omega_{-q})\nonumber \\
 & + & n_{q}\delta(\hbar\omega+\epsilon_{k}-\epsilon_{k+q}+\hbar\omega_{q})]_{\omega\rightarrow0}.\label{eq:mm l_sp}\end{eqnarray}

\noindent with the first term representing a phonon emission and the
second, phonon absorption. $f_{k}=1/(1+e^{\beta\epsilon_{k}})$, $n_{q}=1/(e^{\beta\hbar\omega_{q}}-1)$
are the fermion and boson occupation factors respectively ($\beta=1/k_{B}T$).

It is clear from this formulation that low energy acoustic phonons
$\omega_{q}\sim cq$ ($c$ the sound velocity) cannot contribute to
the scattering because the energy-momentum conservation laws cannot
be simultaneously satisfied. We thus consider scattering by optical
phonons of frequency $\omega_{0}$, as the effect of scattering by
zone boundary acoustic phonons (Umklapp scattering) is similar; we
take a typical $J\sim2400\,\mathrm{K}$. In the fits shown in Fig.~\ref{fig:l fit of Sr2CuO3 with MM}
and Fig.~\ref{fig:l fit of SrCuO2 with MM} in the temperature range
$30\,\mathrm{K}<T<300\,\mathrm{K}$ we assume $l_{\mathrm{mag}}^{-1}\left(T\right)=l_{0}^{-1}+\tilde{l}_{\mathrm{sp}}^{-1}\left(T\right)$
and optimize with respect to $l_{0}$ (the impurity scattering length)
and $\omega_{0}$. As we do not know $\lambda$, we normalize the
theoretical curve with respect to the experimental one at $T=300\,\mathrm{K}$.
For Sr$_{2}$CuO$_{3}$ (SrCuO$_{2}$) a good fit for the whole temperature
range is found with  $l_{0}=1266$ ($l_{0}=3831$) chain unit cells
and $\omega_{0}=540\,\mathrm{K}$ for both compounds. A slightly
better agreement at high temperatures can be achieved with a larger
$\omega_{0}$, although this leads to a strong deviation at low temperatures.
The discrepancy at around 100\,K is due to the fact that the model
does not accurately reproduce the slope of the experimentally determined
$l_{\mathrm{sp}}$ values below this temperature. This is illustrated
in Fig.~\ref{fig:l sp comparison - Data vs MM}. There, an $l_{\mathrm{sp}}$
approximated from the measured data is compared to the $\tilde{l}_{\mathrm{sp}}$
as obtained from the memory function approach. The approximation of
$l_{\mathrm{sp}}$ is done by $l_{\mathrm{sp}}^{-1}\left(T\right)=l_{\mathrm{mag}}^{-1}\left(T\right)-l_{0}^{-1}$,
where an $l_{\mathrm{mag}}$ from equation~\ref{eq:l mag data} and
the estimated $l_{0}=1.56\pm0.16$~\textmu{}m by the phenomenological
model is used. While the agreement between $l_{\mathrm{sp}}$ and
$\tilde{l}_{\mathrm{sp}}$ is good at high temperatures, it gets poorer
towards low temperatures. Especially in the double-logarithmic plot
it can be seen that below 100\,K the model does not reproduce the
slope of the experimental estimate. It is interesting to note, that
the fitted $\omega_{0}$ is somewhat lower but still of the same order
of magnitude than typical frequencies of the optical Cu-O stretching
mode\cite{Lee2000,Popovic2001,Gruninger2000,Windt01}. In contrast,
the phenomenological $l_{\mathrm{sp}}$ according to equation~\ref{eq:general Umklapp}
hints at a scattering by acoustical phonons as can be seen by the
$T_{u}^{*}=210\,\mathrm{K}$.

{We mention that the exponential decay of the heat conductivity parallel to the spin chains $\kappa_{b}\approx\kappa_{\mathrm{mag}}\propto\exp(T_{u}^{*}/T)$ is  consistent with a theoretical treatment by Shimshoni et al.\cite{Shimshoni2003,Shimshoni2005}. However, we do not observe $\kappa_{c}=\kappa_{\mathrm{ph}}\propto\exp(2T_{u}^{*}/T)$ as is expected in the same model. }
\section{Summary}

We have investigated the spinon thermal conductivity $\kappa_{\mathrm{mag}}$
of high-purity single crystals of the single-chain $S=1/2$ AFM Heisenberg
chain compound Sr$_{2}$CuO$_{3}$. We find that $\kappa_{\mathrm{mag}}$
is strongly enhanced as compared to previous results obtained on lower
purity crystals. The analysis of the data yields a very large low-temperature
mean free path of $\sim0.5\text{\textmu m}$, corresponding to 1266
chain unit cells. Upon increasing the temperature towards room temperature,
the mean free path decreases strongly and approaches that observed
in lower purity samples. By using a kinetic model we can decompose
the mean free path into a temperature-independent spinon-defect scattering
length $l_{0}$ and a temperature dependent spinon-phonon scattering
length $l_{\mathrm{sp}}\sim\frac{1}{T}\exp{T_{u}^{*}/T}$ with a characteristic
energy scale $k_{B}T_{u}^{*}$ for umklapp processes. and for low-$T$
upon increasing the purity.

By comparing the temperature dependence of the mean free path of Sr$_{2}$CuO$_{3}$
with that of SrCuO$_{2}$, we could show that the spin-phonon interaction,
as expressed by $l_{\mathrm{sp}}$ is practically the same in both
systems. The comparison of the empirically derived $l_{\mathrm{sp}}$
with model calculations for the spin-phonon interaction of the $S=1/2$
AFM $XY$ chain model yields a reasonable agreement. This agreement
is very encouraging for further studies as an analysis of the full
Heisenberg model might improve this agreement even more. 
\begin{acknowledgments}
We thank W. Brenig, A.~L.~Chernyshev, F.~Heidrich-Meisner, and P. Prelov\v{s}ek
for fruitful discussions. Additionally,
we thank W. Brenig for an important comment on the manuscript. This work was supported by the Deutsche
Forschungsgemeinschaft through grant HE3439/7, through the Forschergruppe
FOR912 (grant HE3439/8) and by the European Commission through the
projects NOVMAG (FP6-032980) and LOTHERM (PITN-GA-2009-238475).
\end{acknowledgments}

\section*{Appendix}

\begin{table}
\begin{centering}
\begin{tabular}{c|c|c|c|c}
 & $B$ in $10^{-31}\,\mathrm{K}^{-1}\mathrm{s}^{2}$ & $A$ in $10^{-43}\,\mathrm{s^{3}}$ & $L$ in $10^{-4}$m & $b$\tabularnewline
\hline 
2N & $2.25$ & $4.50$ & $7.50$ & $3.60$\tabularnewline
\hline 
4N & $3.45$ & $1.53$ & $2.33$ & $3.49$\tabularnewline
\end{tabular}
\par\end{centering}

\caption{Fit parameters of a Callaway fit to $\kappa_{c}$ of Sr$_{2}$CuO$_{3}$,
shown in the left panel of Fig.~\ref{fig:kappa of Sr2CuO3} with
$\alpha=3$ and $\beta=1$. \label{tab:Callaway-Fit parameters}}

\end{table}

In order to model the phononic thermal conductivity perpendicular
to the chain, a phenomenological model, devised by Callaway~\cite{Callaway1959},
can be used. Although this model has undergone several revisions and
extensions~\cite{Callaway1960,Callaway1961,Simons1972,Callaway1991,Sood1993,Chung2004},
the main approach is to model $\kappa_{\mathrm{ph}}$ within the Debye
approximation as~\cite{Berman1976} \begin{equation}
\kappa_{\mathrm{ph}}=\frac{k_{B}}{2\pi^{2}v_{\mathrm{ph}}}\left(\frac{k_{B}T}{\hbar}\right)^{3}\int_{0}^{\Theta_{D}/T}\frac{x^{4}\mathrm{e}^{x}}{\left(\mathrm{e}^{x}-1\right)^{2}}\cdot\tau_{c}\mathrm{d}x.\end{equation}
Here $x=\hbar\omega/k_{B}T$, $\omega$ is the phonon angular frequency,
$\Theta_{D}$ is the Debye temperature and $v_{\mathrm{ph}}$ is the
phonon velocity. In the Debye approximation the sound velocity is
given as \begin{equation}
v_{s}=\frac{\theta_{D}k_{B}}{\left(6\pi^{2}N\right)^{3}\hbar},\label{eq:Sound velocity in Debye approximation}\end{equation}
with $N$ representing the number of elementary cells per unit volume.
$\tau_{c}$ is a combined scattering rate, which is assumed to be
the sum of all individual scattering rates \begin{equation}
\tau_{c}^{-1}=\tau_{B}^{-1}+\tau_{D}^{-1}+\tau_{U}^{-1},\end{equation}
\textcolor{black}{where $\tau_{B}$ denotes boundary scattering, $\tau_{D}$
point defect scattering and $\tau_{U}$ Umklapp scattering. This separation
is possible, as long as these scattering processes are independent
of each other, which is the gist of Matthiessen's rule. In the context
of the model, this is considered to be fulfilled since in different
temperature regions different scattering mechanisms are dominant.
A $\tau_{c}$ containing expressions for all scattering processes
can then be written as} \begin{equation}
\tau_{c}^{-1}=\frac{v_{\mathrm{ph}}}{L}+A\omega^{4}+B\omega^{2}T\exp\left(-\frac{\Theta_{D}}{bT}\right),\end{equation}
with fit parameters $L,$ $A,$ $B$, $b$. The values of these parameters
are given in table~\ref{tab:Callaway-Fit parameters} for the fit
in the inset of Fig.~\ref{fig:kappa of Sr2CuO3}. The largest changes,
when comparing the two purities, are for parameter $A$, which describes
the concentration of point defects, and for parameter $L$, which
describes the boundary scattering. The difference in boundary scattering
only indicates a difference in the sample geometries between the
2N and 4N samples. The decrease of scattering by point defects upon
an increase of purity however underpins the overall reduction of defects.
It should be noted that in Ref.~\onlinecite{Sologubenko2001} two
additional scattering processes had to be used to describe the data.
In our analysis which is focused on temperatures above the maximum
of $\kappa$ these are not necessary. 

\bibliographystyle{BibTex/my-apsrev}
\bibliography{BibTex/Cuprates,BibTex/TiOX,BibTex/ThermalConductivity}

\end{document}